\PassOptionsToPackage{table,svgnames}{xcolor}
\documentclass[conference]{IEEEtran}
\IEEEoverridecommandlockouts
\usepackage[utf8]{inputenc} % input encoding for interpreter
\usepackage[T1]{fontenc}
\usepackage{newtxtext}
\usepackage{multicol}
\usepackage{colortbl} % colors library.
\usepackage[hyphens,spaces,obeyspaces]{url}
\usepackage{cite}
\usepackage{amsmath,amssymb,amsfonts}
\usepackage{subcaption}
\usepackage{algpseudocode}
\usepackage{bmpsize}
\usepackage{graphicx}
\usepackage{tikz}
\graphicspath{ {./images/} }
\usepackage{textcomp}
\usepackage{xcolor}
\usepackage{makecell}
\usepackage[linesnumbered,ruled,norelsize]{algorithm2e}
\makeatletter
\newcommand{\removelatexerror}{\let\@latex@error\@gobble}
\makeatother

\usepackage{mwe}

\SetCommentSty{mycommfont}

\def\BibTeX{{\rm B\kern-.05em{\sc i\kern-.025em b}\kern-.08em
   T\kern-.1667em\lower.7ex\hbox{E}\kern-.125emX}}

\begin{document}

%\title{Cross-Layer Optimization of Big Data Transfers via Decision Tree Based Uncertainty Quantification}
\title{Energy-Efficient Data Transfer Optimization via Decision-Tree Based Uncertainty Reduction}

\author{\IEEEauthorblockN{Hasibul Jamil, Lavone Rodolph, Jacob Goldverg  and Tevfik Kosar} \IEEEauthorblockA{Department of Computer Science and Engineering \\University at Buffalo (SUNY), Amherst, NY 14260, USA} \IEEEauthorblockA{
Email: \{mdhasibu, lrodolph, jacobgol, tkosar\}@buffalo.edu}
}

\maketitle

\begin{abstract}
The increase and rapid growth of data produced by scientific instruments, the Internet of Things (IoT), and social media is causing data transfer performance and resource consumption to garner much attention in the research community. The network infrastructure and end systems that enable this extensive data movement use a substantial amount of electricity, measured in terawatt-hours per year. Managing energy consumption within the core networking infrastructure is an active research area, but there is a limited amount of work on reducing power consumption at the end systems during active data transfers. This paper presents a novel two-phase dynamic throughput and energy optimization model that utilizes an offline decision-search-tree based clustering technique to encapsulate and categorize historical data transfer log information and an online search optimization algorithm to find the best application and kernel layer parameter combination to maximize the achieved data transfer throughput while minimizing the energy consumption. Our model also incorporates an ensemble method to reduce aleatoric uncertainty in finding optimal application and kernel layer parameters during the offline analysis phase. The experimental evaluation results show that our decision-tree based model outperforms the state-of-the-art solutions in this area by achieving 117\% higher throughput on average and also consuming 19\% less energy at the end systems during active data transfers.
%Also augmenting data transfer throughput over high-speed, long-distance networks is becoming increasingly challenging due to numerous factors such as fluctuating network conditions, limitations of underlying transfer protocols, dataset characteristics, network characteristics, and data transfer parameter settings. 
%A multifaceted approach for energy-efficient data transfer optimization involves optimally and dynamically adjusting application-layer and kernel-layer parameters based on real-time network conditions. 
\end{abstract}


\section{Introduction}

The data requirements of commercial and scientific applications continue to increase at an unprecedented rate, generating more data in a given year than the total amount of data generated in the previous years. The advancements ranging from microscale sensor technologies to macroscale supercomputers and large scientific instruments are propelling multi-institutional experimental collaborations between geographically dispersed entities requiring massive data sharing using high-speed wide-area networks. Internet of Things (IoT), social media, and e-commerce applications generate a similar demand in the industry. It is estimated that the number of devices connected to IP networks will exceed 30 billion by 2023, which will be three times the global population~\cite{cisco2020cisco}. As a result, annual data movement across the global Internet has already exceeded the zettabyte scale and continues to increase exponentially. This vast data movement comes with a large economic cost as well as a massive energy footprint. The energy consumption of telecommunication networks has recently exceeded 350 terawatt-hours, and the Internet comprises more than 10\% of the overall energy consumption in many countries, costing the global economy billions of dollars per year~\cite{bell2020}. 

An extensive portion of the existing research on energy-efficient networking focuses on reducing energy consumption in the core networking infrastructure (e.g., switches, hubs, and routers). State-of-the-art power-aware networking techniques include emerging architectures with programmable switches~\cite{greenberg2008towards}, and power-aware networking protocols designed to consider energy consumption while routing data~\cite{chabarek2008power}, and putting idle components to sleep~\cite{gupta2003greening}. Some of these solutions have significant shortcomings. For example, placing the idle components to sleep can be detrimental to performance while saving power. Also, replacing existing network switches and network protocols with power-aware switches and energy-efficient network protocols is a costly solution and not practical in the short term.

In this paper, we introduce a novel decision-tree based cross-layer optimization solution, which is low cost, very easy and practical to deploy, and does not penalize the performance while increasing energy efficiency. Our approach can successfully co-tune several application-layer and kernel-layer parameters to achieve high data transfer throughput while minimizing energy consumption at the same time. 
Our model consists of two phases. In the first phase, an offline
decision-search-tree based clustering technique is used to encapsulate
and categorize historical data transfer log information, and partition the data space from historical data transfer log files that contain network and dataset characteristics~\cite{castin2018}. This step also considers the uncertainty introduced by the stochastic nature of background traffic during a transfer. The goal of the offline analysis is to find the intrinsic structure of historical logs by organizing data objects into similarity groups or clusters~\cite{Liu2004ClusteringVD}. 
In the second phase, an online dynamic search optimization algorithm is used to find the best application and kernel layer parameter combination based on our offline discovered knowledge and the real-time network conditions to maximize the achieved transfer throughput and minimize energy. 

The major contributions of this paper include: 
\begin{itemize}
    \item To the best of our knowledge, our novel multifaceted approach is the first to dynamically construct decision trees based on historical data transfer logs encapsulating network conditions and file characteristics while considering uncertainty introduced from historical data.
    \item Our approach is the first to tune both application-layer and kernel-layer data transfer parameters utilizing novel decision-search-tree based optimization to maximize the achieved data transfer throughput and reduce energy consumption in real-time. \item Our decision-tree based optimization algorithm achieves on average 117\% higher transfer throughput compared to state-of-the-art solutions and 19\% less energy usage in end systems during transfer.
\end{itemize}

% compared to Di Tacchio et al \cite{di2019cross} for 8 out of 9 cases. 
The rest of the paper is organized as follows: Section \textrm{II} presents the problem formulation; Section \textrm{III} discusses our proposed decision-search-tree model and its construction; Section \textrm{IV} presents the evaluation of our model; Section \textrm{V} describes the related work in this field; and Section VI concludes the paper.

\section{Problem Formulation}
In this work, we perform joint tuning of three application-layer parameters (i.e., concurrency, parallelism, and pipelining) and two kernel-layer parameters (i.e., number of active CPU cores and CPU frequency level) to achieve high data transfer performance and low energy consumption at the same time.
Concurrency ($cc$) controls the number of server processes/threads, and these processes/threads can transfer various files independently. As a result, concurrency can accelerate the transfer throughput when many files need to be moved~\cite{liu2010}\cite{kosar2005}\cite{Esma2012}. Parallelism ($p$) is the number of data connections that each server process can open to transfer different portions of the same file in parallel. Parallelism is a fine option for medium or large file transfers in maximizing transfer throughput~\cite{hacker2005}\cite{esma2008}\cite{esma2009}. Pipelining ($pp$) is useful for small file transfers as it eliminates the delay imposed by the acknowledgment of the previous file transfer completion before starting the following one~\cite{effectTCPvariant}\cite{fred2000}\cite{kim-2012}. The kernel-layer parameters, such as the number of active CPU cores ($cpu\_num$) involved in the data transfer and the frequency level of each active CPU core ($cpu\_freq$), also have a significant impact on the achieved transfer throughput and energy consumption.

Given a source endpoint $e_s$ and destination endpoint $e_d$;
a connection link bandwidth $b$ and round-trip-time $rtt$, a dataset with a total size $d_{all}$, average file size $f_{avg}$, buffer size $buf_{size}$, number of files $n$, and the load from contending transfers $l_{ctd}$, the achieved throughput $T$ and  end-system energy $E$
consumption can be expressed as following functions:
% \vspace{-3mm}
\begin{equation}
    T=f_1\left( b, r t t, f_{a v g}, buf_{size}, n, c c, p, p p, cpu_{num},cpu_{freq},l_{c t d}\right)
\end{equation}
\begin{equation}
    E=f_2\left( b, r t t, f_{a v g}, buf_{size}, n, c c, p, p p, cpu_{num},cpu_{freq},l_{c t d}\right)
\end{equation}
For fixed $b, r t t, f_{a v g}, buf_{size},n$, Equation 1 and 2 could be converted to following functions:
\begin{equation}
    T=g_1\left(  \theta,l_{c t d}\right) 
\end{equation}
\begin{equation}
    E=g_2\left(  \theta,l_{c t d}\right)
\end{equation}
where $\theta$ = $\{cc,p,pp,cpu_{num},cpu_{freq}\}$ is the list of tunable  parameters. 
The optimization model is expressed with two types of service-level agreement (SLA) based objective functions: (1) minimum energy or energy-constrained SLA and (2) maximum throughput or throughput guarantee SLA. According to the SLA type and the SLA specifications, we need to find the optimal combination of the tunable parameters $\theta$ to satisfy the SLA requirements. The energy-constrained optimization model determines the matching node from the decision search tree and constructs the "energy surface" containing the matching node. We then find the minimum energy corresponding to the tunable parameters $\theta$ from the energy surface. Similarly, we construct the "throughput surface" containing the matching node from the logs for throughput optimization. We finally find the maximum throughput corresponding to tunable parameters $\theta$ from the throughput surface. The throughput and energy SLA based optimization models are expressed as below:  

\begin{equation}
%$$
\begin{array}{ll}
\underset{\{c c, p, p p, cpu_{num},cpu_{freq} \}}{\operatorname{argmax}} & \int_{t_{s}}^{t_{f}} t h \\
\text { subject to. } & c c \times p \leq \mathbb{N}_{\text {streams }} \\
& p p \leq \mathbb{P} \\
& t h \leq b \\
&E \leq E_{sla}
\end{array}
%$$
\end{equation}

\begin{equation}
%$$
\begin{array}{ll}
\underset{\{c c, p, p p, cpu_{num},cpu_{freq} \}}{\operatorname{minimum}} & \int_{t_{s}}^{t_{f}} e \\
\text { subject to. } 
&T \geq T_{sla}
\end{array}
%$$
\end{equation}

%\vspace{-1mm}
where $t_s$ and $t_f$ are the transfer start and end times respectively. Additionally, $\mathbb{N}_{\text {$streams$}}$ and $P$ are the maximum allowable parameter values in the network. Our goal is to solve the maximum throughput SLA optimization model presented in Equation 5 and the minimum energy optimization model presented in Equation 6. As throughput is achieved in a shared environment, other concurrent contending transfers using the same network resources could affect the behavior of achievable throughput. Our historical logs contain information on such transfers. We define the load from those contending transfers as $l_{c t d}$. There are several assumptions made to create our optimization model, which we describe below:

\begin{itemize}
\item {\em Assumption 1:} All competing transfers at a given time can achieve aggregate throughput, $T=\sum_{i=1}^{N} t h_{i}$, where $N$ is the number of $\mathrm{TCP}$ streams for all the competing transfers, and $t h_{i}$ is the throughput of an individual transfer $i$.
\item {\em Assumption 2:} The fluctuation on transfer behavior (i.e., throughput) depends on all other contending transfers $l_{c t d}$ for given $\theta$=${cc,p,pp,cpu_{num},cpu_{freq}}$ where $b, r t t, f_{a v g}, buf_{size},n$ are fixed.
\item {\em Assumption 3:} Maximum achievable throughput is limited by the end-to-end link bandwidth, disk read speed at the source, or disk write speed at the destination. Given disk read speed $v_{\text {read }}$, disk write speed $v_{\text {write }}$, and the link bandwidth $b$, the maximally achievable end-to-end throughput $t h_{\max }$ would be:
\begin{equation}
%$$
t h_{\max } \leqslant \min \left\{b, v_{w r i t e}, v_{\text {read }}\right\}
%$$
\end{equation}

\item {\em Assumption 4:} Introduced model is agnostic of the end nodes' underlying file systems. Because of parallelism and concurrency, the introduced model would fit superior performance when parallel file systems are used at the end nodes. Performance degradation due to hardware configuration error, storage access delay, and intermediate network system bottlenecks could limit the achievable throughput. Eliminating such bottlenecks might increase the limit of achievable throughput.
\end{itemize}

We chose to accumulate and store real-world historical log data transfer information; and utilize a decision search tree on the historical log files to cluster and group similar log entries based on network conditions, dataset meta-information, and network characteristics. The purpose of this grouping technique is to generalize relationships between throughput and $\theta,l_{c t d}$ as depicted in Equations 3 and 4. The obtained groups of logs are then used for throughput and energy surface construction and these surfaces are consulted to find optimum application and kernel parameter settings to achieve different SLA given $ b, r t t, f_{a v g}, buf_{size},n$ are available. Finally, given a set of attribute parameters $ b, r t t, f_{a v g}, buf_{size},n$ and a SLA, finding the max throughput associated or min energy associated tunable parameters $\theta$=${cc,p,pp,cpu_{num},cpu_{freq}}$ is the objective of this work.

% The target tree data structure will be built from historical data transfer logs in such a way that the optimum application level parameters could be found quickly with different file size and network states ($ b, r t t, f_{a v g}, buff_{size},n$) as attributes.
% % \textbf{One drawback} of this approach is the search key parameter values have to be present in the historical logs. 
% For unknown values of the key parameters, the search result will result in some intermediate non-leaf node contained applications parameters. This process is illustrated with an example in section \textrm{III}(C).

\begin{table*}[h!]
\label{tab:logs}
 \centering
 \caption{Historical transfer log sample with tunable parameters and corresponding achieved throughput and energy expenditure}
\rowcolors{2}{}{gray!10}
\begin{tabular}{|l|l|l|l|l|l|l|l|l|l|l|l|l|}
\hline
\centering
\begin{tabular}[c]{@{}l@{}}\textbf{Entry}\\ \textbf{No}\end{tabular} & \begin{tabular}[c]{@{}l@{}}\textbf{File}\\ \textbf{Size (kB)}\end{tabular} & \begin{tabular}[c]{@{}l@{}}\textbf{\#of}\\ \textbf{Files}\end{tabular} & \textbf{RTT(ms)} & \begin{tabular}[c]{@{}l@{}}\textbf{TCP} \\ \textbf{Buffer}\\ \textbf{Size(MB)}\end{tabular} & \begin{tabular}[c]{@{}l@{}}\textbf{Bandwidth} \\ \textbf{(Mbps)} \end{tabular} & \begin{tabular}[c]{@{}l@{}}\textbf{Throughput} \\ \textbf{(Mbps)} \end{tabular} & \begin{tabular}[c]{@{}l@{}}\textbf{Transfer} \\ \textbf{Energy(J)}\end{tabular} & \textbf{C} & \textbf{P} & \textbf{PP} & \begin{tabular}[c]{@{}l@{}}\textbf{\#of}\\ \textbf{CPU}\end{tabular} & \begin{tabular}[c]{@{}l@{}}\textbf{CPU} \\ \textbf{Frequency(Hz)}\end{tabular} \\ \hline \hline

1                                                  & 100                                                 & 250                                                  & 10  & 200                                                          & 10        & 5          & 20                                                         & 1 & 2 & 2  & 2                                                  & 1.3                                                      \\ 
2                                                  & 100                                                 & 200                                                  & 8   & 150                                                          & 15        & 3          & 17                                                         & 1 & 1 & 1  & 2                                                  & 1.3                                                      \\ 
3                                                  & 50                                                  & 150                                                  & 15  & 250                                                          & 20        & 4          & 15                                                         & 1 & 2 & 1  & 2                                                  & 1.3                                                      \\ 
4                                                  & 40                                                  & 150                                                  & 20  & 225                                                          & 5         & 1          & 12                                                         & 1 & 2 & 2  & 2                                                  & 1.3                                                      \\ 
5                                                  & 150                                                 & 225                                                  & 15  & 150                                                          & 8         & 5          & 22                                                         & 2 & 3 & 3  & 2                                                  & 1.3                                                      \\ 
6                                                  & 100                                                 & 250                                                  & 10  & 200                                                          & 10        & 8          & 15                                                         & 2 & 3 & 3  & 4                                                  & 1.5                                                      \\ 
7                                                  & 100                                                 & 200                                                  & 8   & 150                                                          & 15        & 10         & 10                                                         & 3 & 4 & 4  & 4                                                  & 1.5                                                      \\ 
8                                                  & 50                                                  & 150                                                  & 15  & 250                                                          & 20        & 8          & 9                                                          & 3 & 1 & 4  & 4                                                  & 1.5                                                      \\
9                                                  & 40                                                  & 150                                                  & 20  & 225                                                          & 5         & 4          & 7                                                          & 3 & 2 & 3  & 4                                                  & 1.5                                                      \\
10                                                 & 150                                                 & 225                                                  & 15  & 150                                                          & 8         & 4          & 16                                                         & 2 & 1 & 3  & 4                                                  & 1.5                                                      \\ \hline
\end{tabular}

\label{tab:logs}
\end{table*}

For example, in the data transfer log sample shown in Table~\ref{tab:logs}, we have 5 different groups of attribute/key parameters and their corresponding throughput and transfer energy with different application and kernel parameters. For each group, the historical logs have same $ b, r t t, f_{a v g}, buff_{size},n$ attributes but different throughput and transfer energy resulting from different tunable parameters $\theta$=${cc,p,pp,cpu_{num},cpu_{freq}}$.

\section{Decision Search Tree approach}
\subsection{How Decision Search Tree Approach Works}
To make the search operation more efficient in the offline analysis phase, we construct a decision-search-tree based data clustering from the historical logs. The decision search tree captures the information found in historical logs by grouping similar logs together. The decision search tree is built so that intermediate nodes in the search tree contain the condition to reach leaf nodes, search tree edges contain the particular condition, and leaf nodes contain the log entries associated with the search attributes. Following figure~\ref{fig:d_tree} shows building such  search trees for an example dataset shown in Table ~\ref{tab:logs}. The root node contains all the logs, and for a different level of the tree, different attributes present in the historical log are used to cut the root node into multiple child nodes. Which attribute to choose while cutting a node is selected by an attribute selection scheme. The resultant tree could have different depths because of the attribution selection variability at different tree-building levels. \vspace{1mm}
In the first tree, as shown in figure~\ref{fig:d_tree}(a),  $"Bandwidth (BW)"$ is chosen as the attribute to cut the root node, and all resultant child nodes achieve leaf threshold condition. As a result, the first tree reaches a tree height of 2. 

\begin{figure}[t]
    \begin{centering}
	\begin{subfigure}[t]{0.5\textwidth}
    	\centering
    	\caption{}%Tree 1 when BW is used
        \includegraphics[keepaspectratio=true,width=75mm]{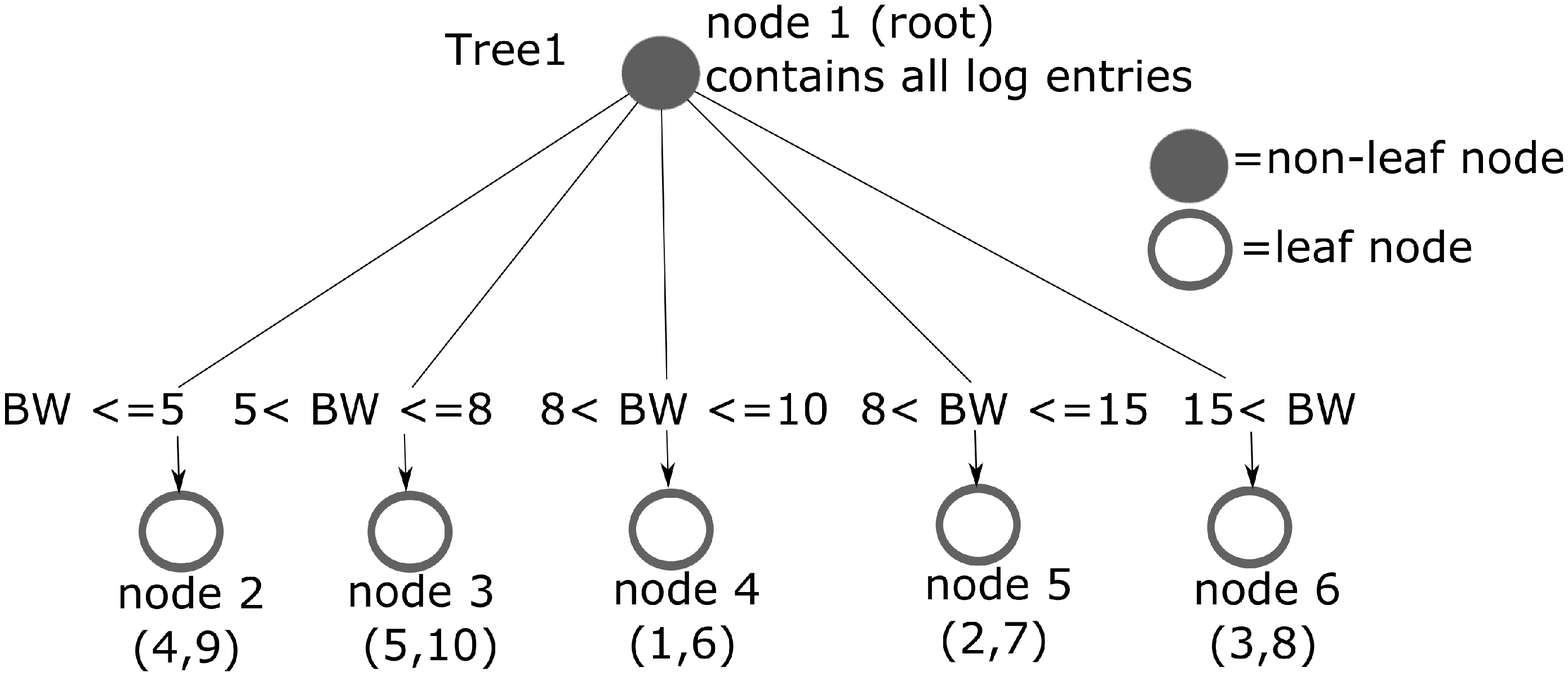}
        \vspace{5mm}
        %\caption{Chameleon Throughput (Mbps)}
    \end{subfigure}
    \begin{subfigure}[t]{0.5\textwidth}
    	\centering
    	\caption{}%Tree 2 when FS and FN are used
        \includegraphics[keepaspectratio=true,width=75mm]{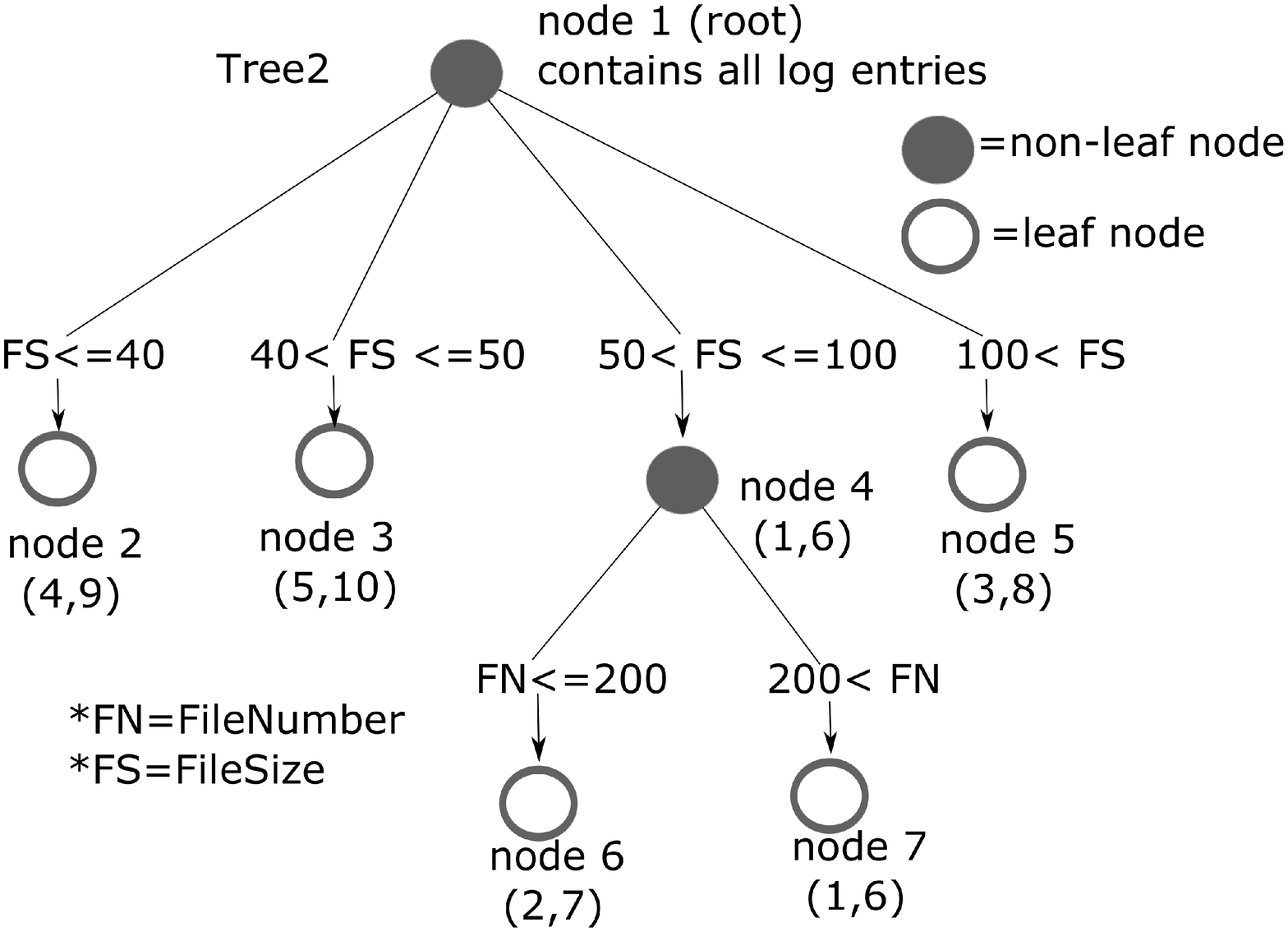}
        %\hspace{-20mm}
        % \vspace{1mm}
        %\caption{CloudLab Throughput (Mbps)}
     \end{subfigure}
     \caption{For Tree1, we are considering bandwidth (BW) as attributes to split nodes and  build complete tree from Table~\ref{tab:logs}. For Tree2, we consider FileSize (FS) and NumberOfFiles(FN) as attributes to split nodes and  build complete tree from Table~\ref{tab:logs}. Resultant trees are different in height based on which attribute set we choose while building the tree. }
     %\vspace{-6mm}
     \label{fig:d_tree}
     \end{centering}
 \end{figure}

In the second tree, as shown in in figure~\ref{fig:d_tree}(b), during the first cut $"FileSize"$ is selected as the attribute to cut the root node, and in the second cut $"FileNumber"$ is chosen to cut node 4, resulting in a tree with depth 3 which is higher than tree height of the previous case. It should be noted that node 4 is the only one that gets cut in level 2 as the other node contained logs small or equal to the leaf threshold. The leaf threshold is the number of minimum logs that makes a node a leaf node, and in this example case, the leaf threshold is equal to 2.

Constructed decision-search-tree based searching for a matching leaf node has a $O(heightOfTree)$ time complexity, and we can build different trees for the same dataset following a different set of attributes. The generated trees, in that case, will be different in height. For a given set of attributes ($f_{avg},n,rtt,buf_{size},b$), to find the matching leaf node in the search tree, the average case search time complexity is better than worst-case search time complexity as not all the leaf nodes will be on the same depth of the tree. Algorithm~\ref{alg:decision search tree} is used to build the decision tree. For an incoming data transfer request, the associated attributes are first obtained, and then these attributes are used to traverse the tree to reach the leaf node. The matched leaf nodes containing historical logs are then used to construct throughput and energy surfaces. By analyzing those surfaces, optimal application and kernel parameters are finally obtained. 

\begin{algorithm}
	\SetKwData{Left}{left}\SetKwData{This}{this}\SetKwData{Up}{up}
	\SetKwFunction{Union}{Union}\SetKwFunction{FindCompress}{FindCompress}
	\SetKwInOut{Input}{input}\SetKwInOut{Output}{output}
	\Input{dataset, leaf\_threshold, cutNumber}
	\tcp {leaf\_threshold is minimum number of logs that declare a node as leaf node}
	\Output{decision search tree}
	\BlankLine
	availableNodeToCutList=[root]
	\tcp {start from root node $S_0$ that contains all the logs from Historical log}
	\While{availableNodeToCutList is not empty}{
	    currentNode=Pop(availableNodeToCutList);\\
	    chooseAttributeToCut(currentNode);      \\
	    \tcp {use DI or SD metric to choose which dimension to cut the node}
	    childNodeList=CutNodeOnChoosenAttribute(); \\
	    \tcp {cut the node on the chosen dimension/attribute cutNumber times}
	        \For{each childNode in ChildNodeList}{
	            Assign matched logs to childNode; \\
	            \tcp {some number of logs from parent node will fall into each child node}
	            childNodeLogCount=NumberOfLoginNode(childNode);
	            \If{childNodeLogCount <= $leaf\_threshold$ OR $sameChildNodeAttributeValue$ }{
                    Continue;\\	            
	            }
	            Push ChildNode to availableNodetocutList ;\\
	        }
	}
	\Return {root} \\ 
	\caption{Decision tree construction algorithm}
	\label{alg:decision search tree}
	
\end{algorithm}
%\DecMargin{1em}
% \vspace{-8mm}

%\DecMargin{1em}
% \vspace{-5mm}
\subsection{Handling of the Unseen Attributes}
The system will return the last matched node as the final node for unseen attributes (i.e., not present in the history log) from an incoming data transfer request. For example, an incoming data transfer request with attributes  $f_{avg},n,rtt,buf_{size},b$ respectively may not match any entries in the history log. For data transfer requests like this, the search will report application and kernel parameters from the constructed throughput surfaces of the last matched node of the tree where it needs not be a leaf node. If the attributes of the data transfer request will result in a match with node $N1$ then node $N1$ corresponding logs, and those logs contained throughput surfaces will be investigated to report final application and kernel parameters $\theta$. For example, key parameter (100,255,10,200,10) (i.e.,$f_{avg},n,rtt,buf_{size},b$)  will match node 4 on second tree shown in III(A) for fileSize attribute value (i.e.,100). The fileNumber attribute value (i.e., 255) is closer to node 7 of tree 2, so it will result in a match with node 7 and node 7 corresponding logs will be used to construct throughput and energy surfaces to find final application and kernel parameters $\theta$. 

% \begin{figure}
%  \label{fig:d_tree}
%   \centering
%   \def\svgwidth{\linewidth}
%     \input{decision_tree.eps_tex}
%   \caption{In a matched leaf node, different throughput surfaces could be constructed for various application and kernel parameters. The figure legends denote five attributes ($ n, f_{a v g}, buff_{size}, b, r t t  $) that construct all the logs on that matched leaf node. This plot aggregates over multiple y-axis values at each value of the x-axis and shows an estimate of the central tendency and a confidence interval for that estimate.}
% % \Description{System architecture for proposed packet classification engine}
% %   \vspace{-5mm}
%   \label{fig:d_tree}
% \end{figure}

% \begin{figure}[t]
%  \label{fig:tree1}
%   \centering
%   \includegraphics[keepaspectratio=true,width=58mm]{images/decision_tree.eps}
%  \vspace{-3mm}
%   \caption{Tree1 Considering "File size and NumberOf Files" as attributes to split nodes and  build complete tree from Table~\ref{tab:logs}}
% % \Description{System architecture for proposed packet classification engine}
%   \vspace{0mm}
%   \label{fig:tree1}
% \end{figure}

% \begin{figure}[t]
%  \label{fig:tree2}
%   \centering
%   \includegraphics[width=\linewidth]{Slide2.png}
%   \caption{Tree2 Considering "Bandwidth" as attributes to split nodes and build tree from Table~\ref{tab:logs}}
% % \Description{System architecture for proposed packet classification engine}
%   \vspace{0mm}
%   \label{fig:tree2}
% \end{figure}

\subsection{Dealing with Uncertainty of Search Result}
Uncertainty could be defined as situations consisting of unknown or partial knowledge and this type of uncertainty is fused in any stochastic and partially observable environments~\cite{ayyub2006}. Epistemic uncertainty captures uncertainty in the model parameters and aleatoric uncertainty is the uncertainty caused by the noise that is inherited in the training dataset~\cite{Moloud2020}. The historical log for data transfer suffers partial observation problems as computing clusters and network usually does not give real-time performance metrics for computation or network system resources. For example, storage servers utilization metric and queue size of devices in shared networks are some metrics that play paramount importance in transfer performance but these parameter metrics are private. The lack of these important real-time performance metrics from all components of end-to-end transfers across multiple clusters and network domains makes the historical dataset incomplete and introduces an aleatoric type of uncertainty. 

In our work, the uncertainty is captured in equation 3 and 4 of our model as the contending transfers $l_{c t d}$ parameter is not deterministic, the search result using the decision tree has aleatoric uncertainty associated with it. The decision search tree uncovers relationships between throughput and $\theta,l_{c t d}$ and the level of the aleatoric type of uncertainty could be reduced by combining the search results made by an ensemble of decision trees and performing the throughput or energy surface construction on the combined search result logs.

An ensemble approach uses the combined output of a set of decision trees (e.g., decision search trees in our case) to improve on the search result accuracy offered by any one of its members~\cite{osti_1527311}\cite{MAHJOUR2022109822}. In our example, we build two trees as shown in section III(A) for sample log shown in Table~\ref{tab:logs}. For an incoming data transfer request with attributes parameters as $f_{avg},n,rtt,buf_{size},b$ respectively, the request matches with node $N1$ in first tree and with node $N2$ in second tree. We combine the historical logs from node $N1$ in tree-1 and logs from node $N2$ in tree-2 to construct the combined logs throughput and energy surfaces. Finally, these surfaces are then used to find maximum throughput  or minimum energy points and associated optimum application and kernel parameter settings for that incoming data transfer request.

% \vspace{1mm}
\subsection{Constructing Decision Search Tree}
The attribute selection procedure is novel in the proposed decision search tree building process where we consider diversity index and standard deviation as metrics from the historical log attribute set to choose the attribute at which we perform node cutting while building the tree. 
\subsubsection{Ranking the attributes based on Diversity Index}
While building the search trees, we expect to build trees with minimum depth (i.e., traversing the tree is faster) and group the historical logs so that the grouped logs are as sparse as possible. To meet both of these objectives, choosing the appropriate attribute to cut a tree node is important. For example, we could choose any of the five attributes from the root node to derive the first layer (i.e., depth=1) nodes. The attribute should be chosen so that it maximally differentiates the historical logs. Diversity Index is such a metric that could be used to rank the attributes, and the ordered attributes could then be used to choose one attribute while building the tree \cite{jamil_multibit}. The derivation of $diversity$ $index $ for an attribute from a history log is derived as follows:

If $S$ is a parameter value set of an attribute $S=\left\{c_{1}, c_{2}, c_{3} \cdot \cdots \cdot c_{n}\right\}$ 
and $C_{max}=\max\left\{c_{1}, c_{2}, c_{3} \cdot \cdots \cdot c_{n}\right\}$, 
$S_{normalized}=\left\{\frac{c_{i}}{C_{\text {max}}}\right\}$  for $i$=1 to $i$=n. So, $S_{normalized}=\left\{c_{1-norm}, c_{2-norm}, c_{3-norm} \cdot \cdots \cdot
c_{n-norm}\right\}$ and
$S_{n-no-duplication}=\left\{c_{1-n-nd}, c_{2-n-nd} \cdot \cdots \cdot c_{n-n-nd}\right\}$
where, $S_{n-no-duplication} \subseteq S_{normalized}$ \\
if $C_{max-norm}=\max\left\{c_{1-norm}, c_{2-norm}, c_{3-norm} \cdot \cdot c_{n-norm}\right\}$\\
$C_{min-norm}=\min\left\{c_{1-norm}, c_{2-norm}, c_{3-norm} \cdot \cdot c_{n-norm}\right\}$, \\
then $diversity$  $index $ ($DI$) could be calculated by following equation:

% \vspace{-5mm}
% \begin{equation}
% \label{eq:2}
% DI=(C_{max-norm } - C_{min-norm }) \mathlarger\times \mathlarger\sum_{i=c_{1-n-nd}}^{i=c_{n-n-nd}} \frac{1}{\text{$freq.$ $of$ }C_{i}}
% %\vspace{-5mm}
% \end{equation}

$D I=\left(C_{\text {max }-\text { norm }}-C_{\text {min-norm }}\right) \times \sum_{i=c_{1-n-n d}}^{i=c_{n-n-n d}} \frac{1}{\text { freq. of } C_{i}}$ \\

The first part of above equation captures how much range a particular attribute (for example, $FileSize$) has, and the second part captures how unique the values for that specific attribute are. The second part of equation is constructed in such way to penalize multiple occurrences of a particular value. 

\subsubsection{Ranking the attributes based on Standard Deviation}
Standard deviation is another metric that could be used to rank the attributes, and the ordered attributes could be used to build the tree while cutting a tree node. Standard deviation measures the relative spread of the values for each attribute field. The standard deviation for a particular attribute value is bigger when the differences of those attribute values are more spread out regarding the distribution mean. For a set of $N$ numbers $\left\{x_{1}, x_{2}, \cdots, x_{N}\right\}$, if the mean of the collection is $\bar{x}$ then standard deviation, $SD$ is as  follows:
\begin{equation}
{SD}=\sqrt{\frac{1}{N-1} \sum_{i=1}^{N}\left(x_{i}-\bar{x}\right)^{2}}
\end{equation}
The variance was also considered to measure the spread of the fields in a ruleset, and described as follows:
\begin{equation}
S^{2}=\frac{\sum\left(x_{i}-\bar{x}\right)^{2}}{N-1}
\end{equation}
where $S^{2}=$ sample variance, $x_{i}=$ the value of the one observation, $\bar{x}=$ the mean value of all observations, and $N=$ the number of items in the set.
\subsubsection{Band of trees}
We build two trees for a given historical transfer log; one with diversity index as attribute selection metric and another with standard deviation as selection metric. Once tree-1 and tree-2 are built, the system traverses both trees to find the matching nodes from each tree for an incoming data transfer request with five attributes ($f_{avg},n,rtt,buf_{size},b$). After traversing and matched nodes are found in the respective trees, application and kernel parameters are obtained as described in Section III(C).
% \vspace{-3mm}
\subsection{Searching $\theta$ for Incoming Data Transfer Request}
Given a historical log "$L$", the decision search tree leaf nodes cluster/group the logs based on combination of five attributes ($f_{avg},n,rtt,buf_{size},b$). The clusters are ${c_{1}, c_{2}, c_{3} \cdot \cdots \cdot c_{n}}$ where $n$ is the number of clusters and is equal to the number of leaf nodes in a search tree.
For each incoming transfer request, corresponding five attributes ($f_{avg},n,rtt,buf_{size},b$) along with target throughput $T_{es}$ or target energy $E_{es}$are used to traverse the search trees and find corresponding matching leaf nodes. Algorithm~\ref{alg:findingOptimalParameters} describes the tree traversing and optimal parameter finding steps.

\begin{algorithm}
	\SetKwData{Left}{left}\SetKwData{This}{this}\SetKwData{Up}{up}
	\SetKwFunction{Union}{Union}\SetKwFunction{FindCompress}{FindCompress}
	\SetKwInOut{Input}{input}\SetKwInOut{Output}{output}
	\Input{$\beta$ =Incoming transfer request attributes $ b, r t t, f_{a v g}, buff_{size},n$, decision tree,targetThroughput $T_{es}$ or targetEnergy $E_{es}$ as SLA}
	\tcp {targetThroughput is the Throughput the algorithm is targeting to achieve based on SLA and targetEnergy is the transfer energy budget based on SLA}
	\Output{$\theta$={cc,p,pp,cpu\_{num},cpu\_{freq}} }
	\BlankLine
% 	matchedNode=traverseTree(\beta);\\ 
    allTheNodesOfTreeList=getAllTheNodesAsList(decision tree);\\
    \While{allTheNodesOfTreeList is not empty}{
	    currentNode=Pop(allTheNodesOfTreeList);\\
	    \If {currentNode has $\beta$}{
	        matchedNode=currentNode;\\
	        break;
	        }
	}
	\If {SLA==MAXThroughput}{
	surfaces= decomposedThroughputSurface(matchedNode);\\
	\tcp {matchedNode throughput surface is decomposed into multiple components with different throughput level}
	finalThroughputSurface=getFinalSurface(surfaces,$T_{es}$);\\
	$\theta$=maxSurfacePoint(finalThroughputSurface);\\ 
	}
	\Else {
	surfaces= decomposedEnergySurface(matchedNode);\\
	\tcp {matchedNode energy surface is decomposed into multiple components with different energy level}
	 finalEnergySurface=getFinalSurface(surfaces,$E_{es}$);\\
	 $\theta$=minSurfacePoint(finalEnergySurface);\\
	}
    \Return {$\theta$} \\ 
	\caption{Optimal parameter discovery algorithm}
	\label{alg:findingOptimalParameters}
\end{algorithm}

From a matched node containing historical logs, we construct throughput as a polynomial surface which is a function of the application and kernel parameters $\theta$ =${cc,p,pp,cpu_{num},cpu_{freq}}$. Incoming transfer request corresponding attributes $f_{avg},n,rtt,buf_{size},b$ matches one of the leaf node in the decision search tree and that node corresponding logs could construct multiple surfaces. The matched surface could then be decomposed into multiple components (i.e., binned surface components) based on the different binned values of throughput (i.e., throughput 0-100 could be binned to bin 100, 101-200 to bin 200, and so on) so that expected throughput could be matched with a binned surface component.

%  as shown in Figure~\ref{fig:surfacesNode}
% as described in Algorithm ~\ref{alg:findingOptimalParameters} line 2. 
In line 10, the algorithm finds the closest throughput surface component from expected throughput $T_{es}$. In line 11 and 12 and calculate the maximum throughput point in the selected surface component, and return the maximum throughput point corresponding to application and kernel parameters as final $\theta$=${cc,p,pp,cpu\_{num},cpu\_{freq}}$. Similar is true for finding min energy SLA corresponding $\theta$ as shown in line 15,16 and 17. Using algorithm ~\ref{alg:findingOptimalParameters}, we pre-compute and store optimized kernel-level and application-level transfer parameters in our custom data structure for our dynamic online program to use.

\begin{table}[!t]
\centering
%\caption{table description}
\label{t_sim_4}
\caption{Characteristics of testbeds}
\rowcolors{2}{}{gray!10}
\begin{tabular}{|l|l|l|l|l|}
\hline
\multicolumn{1}{|l|}{\textbf{Testbed}} & \multicolumn{1}{l|}{\textbf{Bandwidth}} & \multicolumn{1}{l|}{\textbf{RTT}} & \multicolumn{1}{l|}{\textbf{BDP}} &
\multicolumn{1}{l|}{\textbf{CPU Architecture}} \\ 
\hline \hline 
Chameleon & 10 Gbps & 34 ms & 40 MB & \makecell{Haswell (server) \\ Haswell (client)} \\  
CloudLab & 1 Gbps & 38 ms & 4.5 MB & \makecell{Haswell (server) \\ Broadwell (client)} \\
Inter-Cloud & 1 Gbps & 45 ms & 4.5 MB & \makecell{Haswell (server) \\ Bloomfield (client)} \\
\hline
\end{tabular}
\end{table}

\subsection{Online Dynamic Maximum Throughput Parameter Tuning}
We cluster historical log entries based on network characteristics, network conditions, and file characteristics during the offline analysis phase. 
% We further perform temporal clustering based on timestamps, RTT, and the corresponding achieved throughput. This is necessary as low throughput could be caused by sub-optimal transfer parameters, high network congestion, or both. We then construct decision search trees based on the rendered clusters. 
Using algorithm~\ref{alg:findingOptimalParameters}, we pre-compute and store optimized kernel-level and application-level transfer parameters for an exhaustive combinations of scenarios in our custom data structure for our dynamic online program to use. Table II shows the offline analysis time for three different testbed with different number of historical logs. The offline analysis consist of two phases: decision tree construction time and decision tree traversing time while generating the custom data structure for our dynamic online program.  Algorithm~\ref{alg:dynamicTuning} shows the steps to use the pre-computed data structure from section III(E) for maximum throughput optimization SLA. Before a data transfer task starts, our online dynamic  tuning algorithm for throughput first measures the RTT for 3 seconds which is shown in line 3 of algorithm ~\ref{alg:dynamicTuning}. Our algorithm utilizes the measured average RTT and the associated expected throughput entry along with other file characteristics as the search key to retrieve initial transfer parameters ($\theta$) from the offline stage pre-computed data structure. Using the initial transfer parameters, the algorithm performs the data transfer for approximately 10 seconds before obtaining new transfer parameters from the decision search tree utilizing the measured average delta RTT and the instantaneous delta throughput. This is necessary to ensure previously retrieved parameters (theta) have not become sub-optimal. Since network conditions may fluctuate in a shared network, we periodically check and adjust transfer parameters for the reasons mentioned above.  We developed three periodic time intervals to check and adjust transfer parameters based on network conditions and dataset characteristics. For small, medium, and large datasets, we review and adjust parameters every 10, 20, and 30 seconds respectively as shown in line 7 to 10 of algorithm~\ref{alg:dynamicTuning}. 
% \iffalse 
% \begin{figure}[!t]
% \removelatexerror% Nullify \@latex@error

\begin{algorithm}
	\SetKwData{Left}{left}\SetKwData{This}{this}\SetKwData{Up}{up}
	\SetKwFunction{Union}{Union}\SetKwFunction{FindCompress}{FindCompress}
	\SetKwInOut{Input}{input}\SetKwInOut{Output}{output}
% 	\Input{Dataset}
% 	\Output{\theta=${cc,p,pp,cpu\_{num},cpu\_{freq}}$ }
	\BlankLine
% 	matchedNode=traverseTree(\beta);\\ 
    datasets=$ClusterFiles()$\\
    $Timeout = getTimeout(Dataset.AvgFileSize)$ \\
    $\delta$ rtt = $measureRtt(time\_3sec)$\\
    $\theta$ = obtainInitParams($\delta$ rtt)\\ 
    $startTransfer$($\theta$, Dataset) \\
    % $initTimeout$ = 10 sec \\
    % \For{$initTimeout$}{
    % $\delta$ Throughput = $measuredInstTrhoughput()$ \\
    % $\theta$ = $SearchTreeParams($\delta$ rtt,$\delta$ Throughput)$ \\
    % $UpdateParameters$($\theta$) \\
    % }
    \For{$Timeout$}{
    $\delta$ throughput = $measuredInstTrhoughput()$ \\
    $\delta$ rtt = $measuredRtt()$ \\
    $\theta$ = SearchTreeParams($\delta$rtt,$\delta$ Throughput) \\
    UpdateParameters($\theta$, Dataset) \\
    %  calculateWeightedThroughput() \\
    %  calculateDeltaRtt()  \\
    %  calculateExternalNetworkPercentage()\\
    %  startWithLightExtNetworkLoadSurface()\\
    }%end For
	\caption{Dynamic Throughput Tuning}
	\label{alg:dynamicTuning}
\end{algorithm}

\subsection{Online Dynamic Minimum Energy Parameter Tuning}
Based on the SLA, if the given SLA is minimum energy, a dynamic online energy constraint tuning algorithm shown in Algorithm~\ref{alg:dynamicEnergyTuning} derived from the offline energy constraint optimization model. Based on the energy constraint SLA, parameter tuner periodically monitors the instantaneous energy consumption at specified regular time intervals as shown in line 7 in Algorithm~\ref{alg:dynamicEnergyTuning}. It then uses this instantaneous energy consumption and transfer elapsed time (i.e., line 8) to approximate the energy consumption of the transfer. With the approximate energy value and current measured delta rtt (i.e., line 3), the dynamic tuner obtains the updated transfer parameters (i.e., line 10). 
% If the instantaneous power consumption exceeds the threshold specified in the SLA, it obtains new optimal parameters within the confidence range/sampling region by obtaining a new energy polynomial surface that is closest to the current mea- sured external network load and energy consumption range. If the measured instantaneous power consumption decreased from the last interval check, it obtains the closest energy polynomial surface and retrieves the associated optimal data transfer parameters.

\begin{algorithm}
	\SetKwData{Left}{left}\SetKwData{This}{this}\SetKwData{Up}{up}
	\SetKwFunction{Union}{Union}\SetKwFunction{FindCompress}{FindCompress}
	\SetKwInOut{Input}{input}\SetKwInOut{Output}{output}
% 	\Input{Dataset}
% 	\Output{\theta=${cc,p,pp,cpu\_{num},cpu\_{freq}}$ }
	\BlankLine
% 	matchedNode=traverseTree(\beta);\\ 
    datasets=$ClusterFiles()$\\
    Timeout = $getTimeout(Dataset.AvgFileSize)$ \\
    $\delta$ rtt = $measureRtt(time\_3sec)$\\
    $\theta$ = $obtainInitParams(\delta rtt)$ \\
    $startTransfer(\theta, Dataset)$ \\
    % $initTimeout$ = 10 sec \\
    % \For{$initTimeout$}{
    % $\delta$ Energy = $measuredInstEnergy()$ \\
    % elapsedTime=$getElapsedTime()$\\
    % approximateEnergy = $enrgyApproximation(\delta Energy, elapsedTime)$ \\
    % $\theta$ = $SearchTreeParams(\delta rtt,approximateEnergy)$ \\
    % $Transfer$($\theta$, Dataset) \\
    % }
    \For{$Timeout$}{
    $\delta$ Energy =$ measuredInstEnergy()$ \\
    elapsedTime=$getElapsedTime()$\\
    approximateEnergy = $enrgyApproximation(\delta Energy, elapsedTime)$ \\
    $\theta$ = $SearchTreeParams(\delta rtt,approximate Energy)$ \\
    $Transfer$($\theta$, Dataset) \\
    %  calculateWeightedThroughput() \\
    %  calculateDeltaRtt()  \\
    %  calculateExternalNetworkPercentage()\\
    %  startWithLightExtNetworkLoadSurface()\\
    }%end For
	\caption{Dynamic Energy Tuning}
	\label{alg:dynamicEnergyTuning}
\end{algorithm}
% \begin{algorithm}
% % \label{alg:dynamicTuning}
%   $\textbf{input:} Dataset $ \\
%   $\textbf{output:} \theta=${cc,p,pp,cpu\_{num},cpu\_{freq}}$ $ \\
%   $Timeout = getTimeout(Dataset Avg. File Size)$ \\
%   $\delta rtt = measureRtt(time\_3sec)$\\
%   $\theta = obtainInitParams(\delta rtt)$ \\
%   $startTransfer(\theta, Dataset)$ \\
%   $initTimeout = 10 sec$ \\
%   \For{$initTimeout$}{
%   $\delta Throughput = measuredInstTrhoughput()$ \\
%   $\theta = SearchTreeParams($\delta rtt,$\delta Throughput)$ \\
%   }
%   \For{$Timeout$}{
%     $\delta throughput = measuredInstTrhoughput()$ \\
%     $\delta rtt = measuredRtt()$ \\
%     $\theta = SearchTreeParams($\delta rtt,$\delta Throughput)$ \\
%     %  calculateWeightedThroughput() \\
%     %  calculateDeltaRtt()  \\
%     %  calculateExternalNetworkPercentage()\\
%     %  startWithLightExtNetworkLoadSurface()\\
    
% }%end For 
% \label{alg:dynamicTuning}
% \caption{Dynamic Online Parameter Tuning}
% \end{algorithm}
% % \end{figure}
% % \fi

\vspace{-1mm}

\begin{figure*}[h]
    \begin{centering}
	\begin{subfigure}[t]{0.31\textwidth}
    	\centering
    	\caption{Chameleon Throughput (Mbps)}
        \includegraphics[keepaspectratio=true,width=58mm]{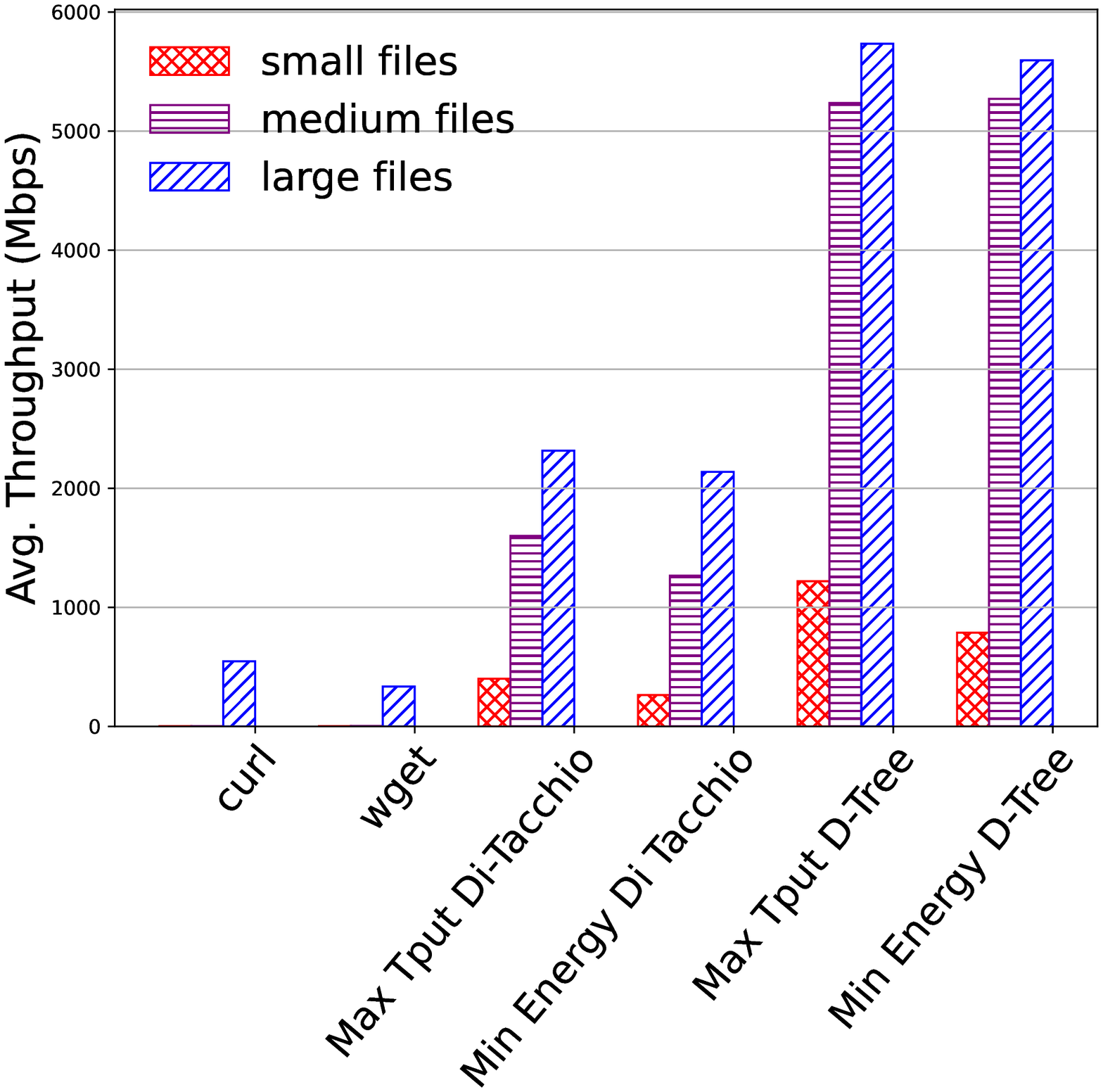}
        \vspace{-2mm}
        %\caption{Chameleon Throughput (Mbps)}
    \end{subfigure}
    \begin{subfigure}[t]{0.31\linewidth}
    	\centering
    	\caption{ CloudLab Throughput (Mbps)}
        \includegraphics[keepaspectratio=true,width=58mm]{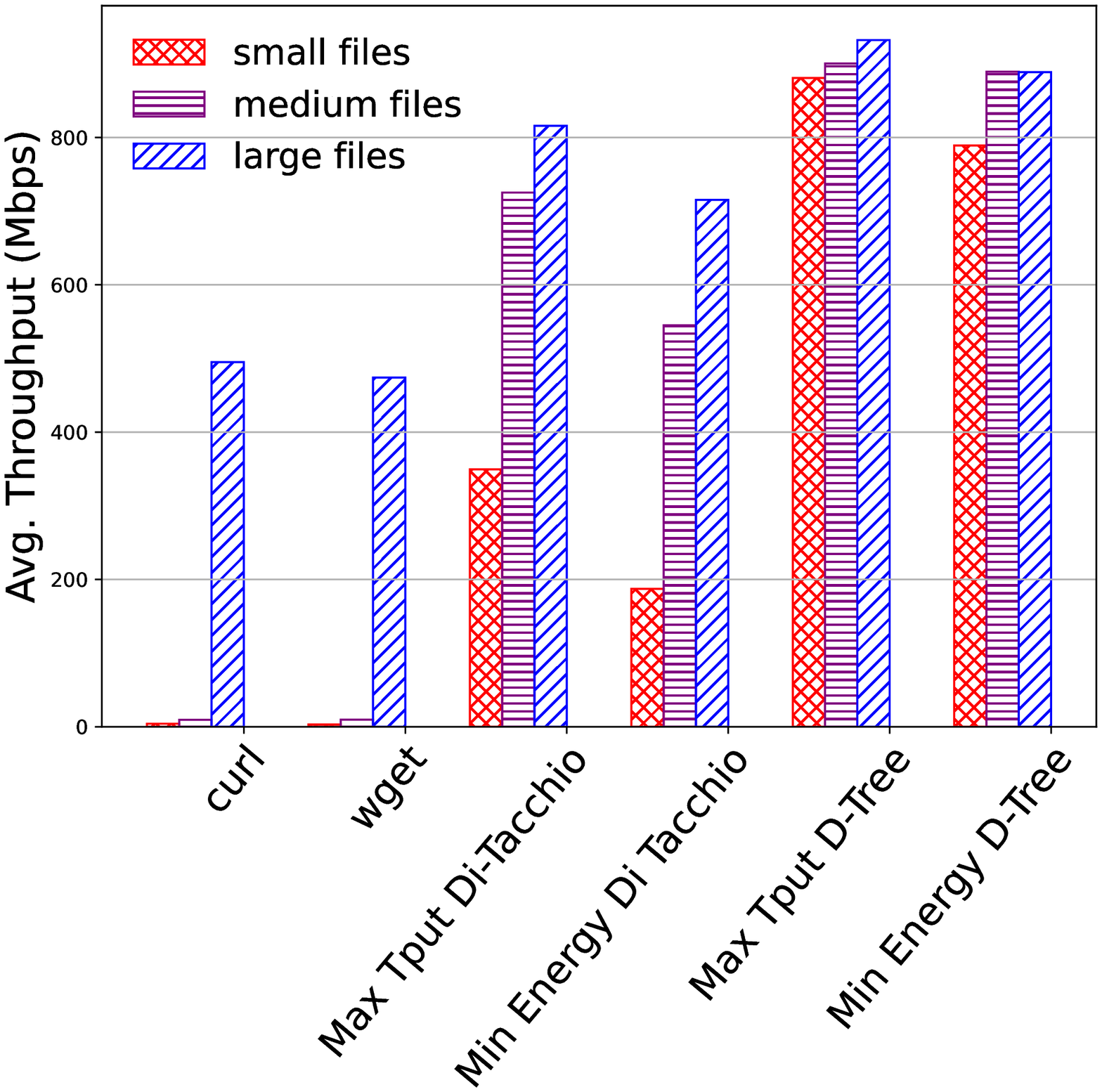}
        %\hspace{-20mm}
        \vspace{-2mm}
        %\caption{CloudLab Throughput (Mbps)}
    \end{subfigure}
    \begin{subfigure}[t]{0.31\textwidth}
    	\centering
    	\caption{Inter-Cloud Throughput (Mbps)}
        \includegraphics[keepaspectratio=true,width=58mm]{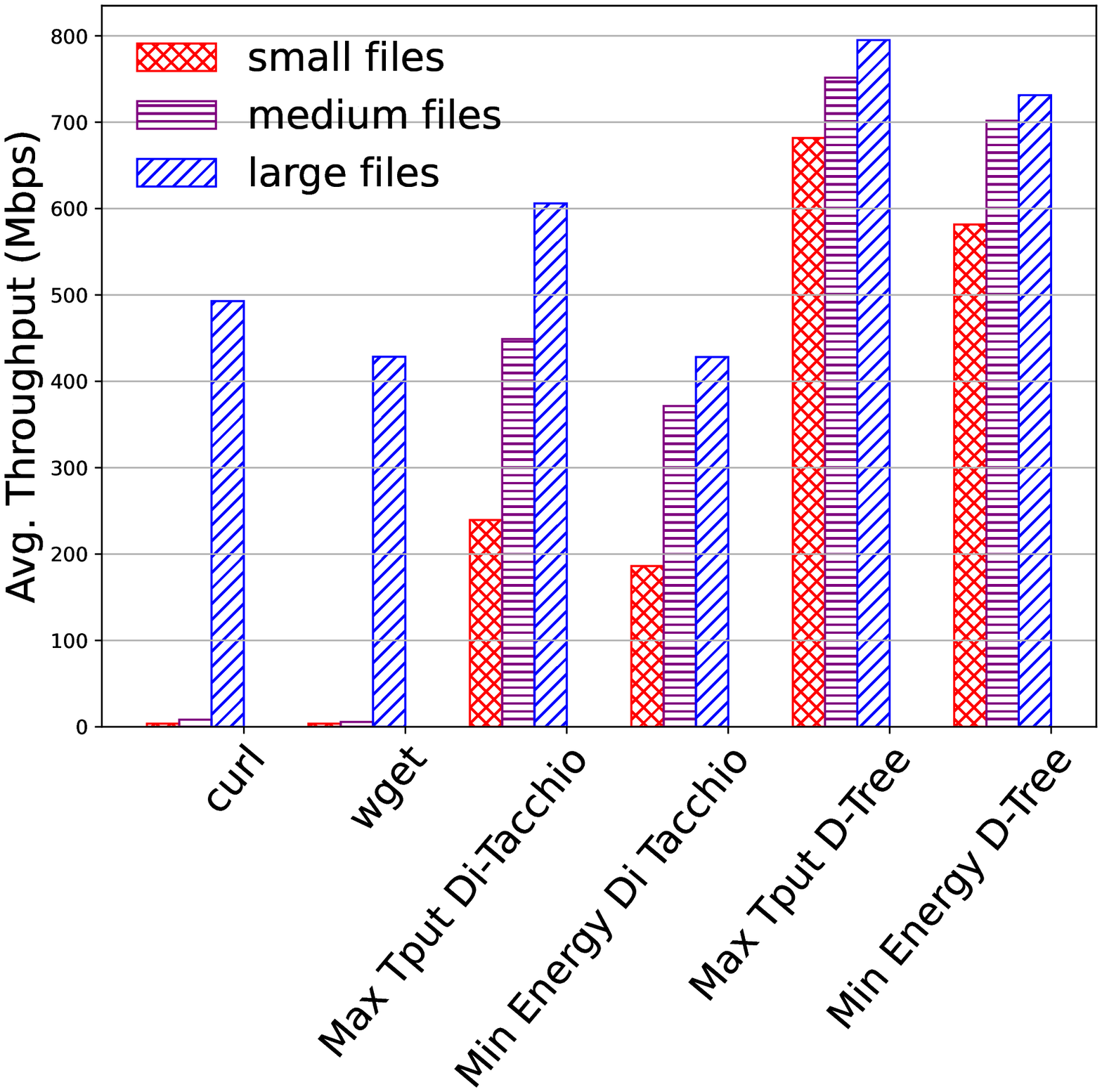}
        \vspace{-2mm}
        %\caption{Inter-Cloud Throughput (Mbps)}
    \end{subfigure}
	\begin{subfigure}[t]{0.31\textwidth}
    	\centering
    	\caption{Chameleon Client Energy (Joules)}
        \includegraphics[keepaspectratio=true,width=58mm]{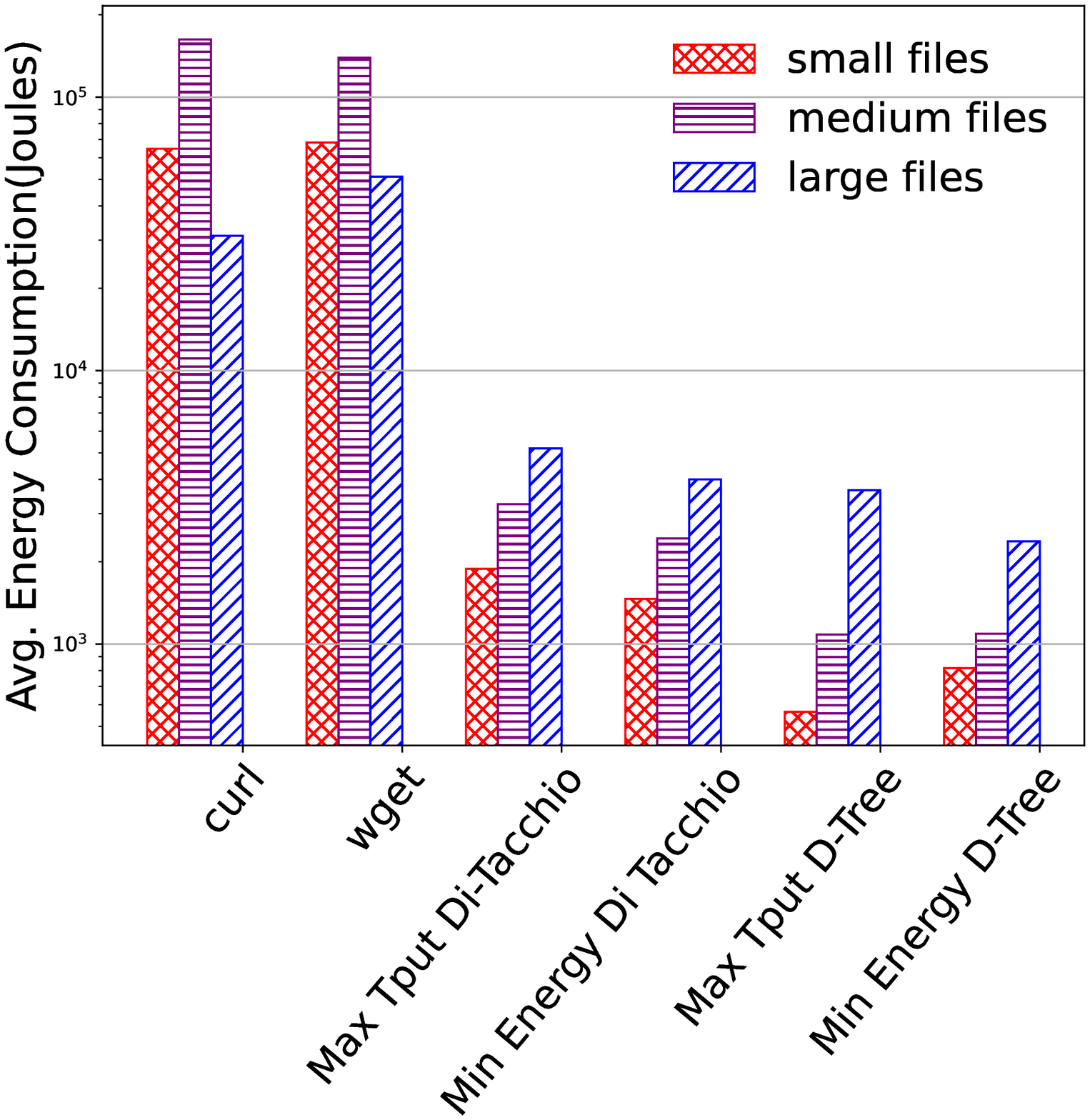}
        \vspace{-2mm}
        %\caption{Chameleon Client Energy (Joules)}
    \end{subfigure}
 	\begin{subfigure}[t]{0.31\textwidth}
    	\centering
    	\caption{Cloulab Client Energy (Joules)}
        \includegraphics[keepaspectratio=true,width=58mm]{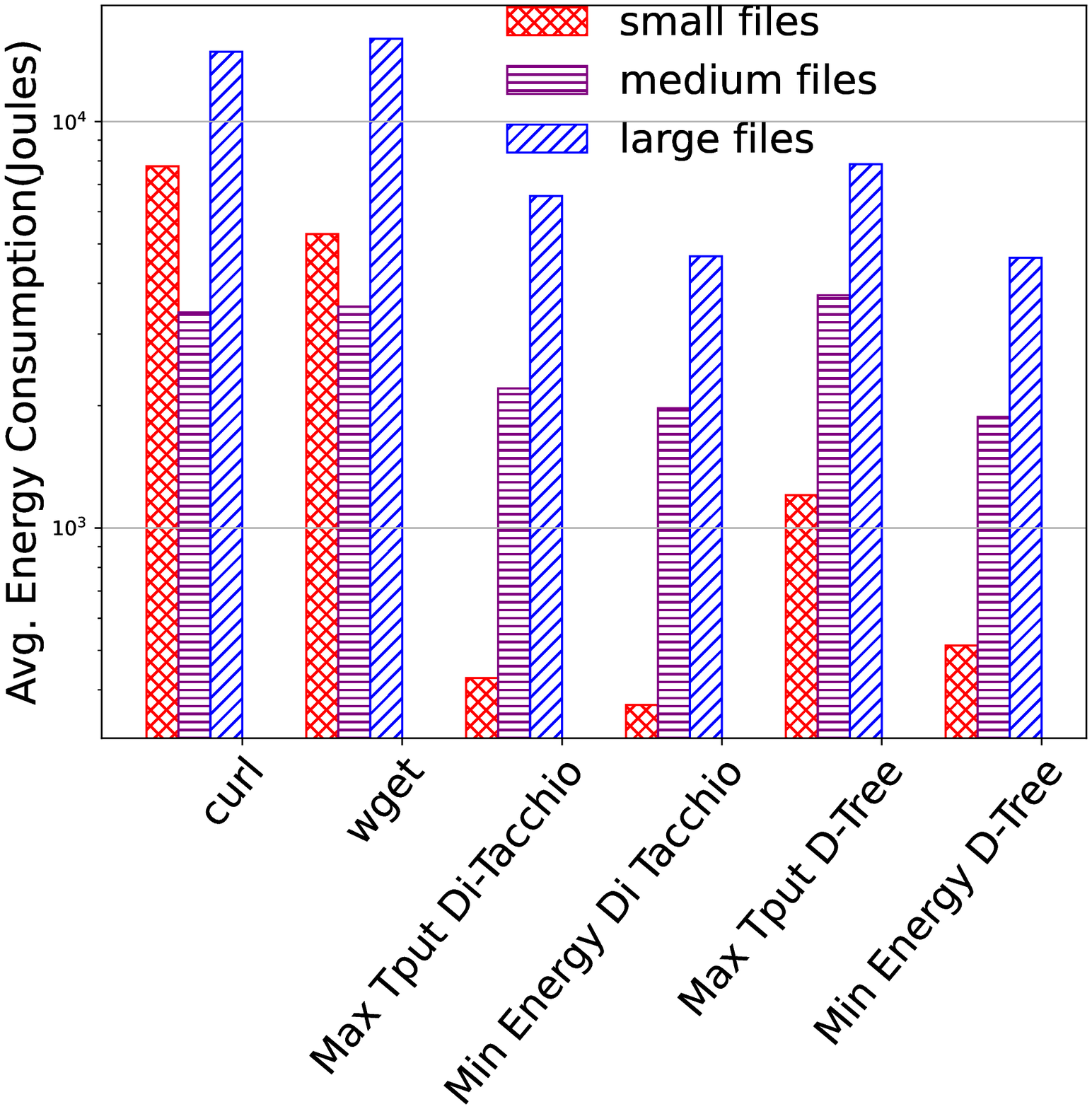}
        \vspace{-2mm}
        %\caption{CloudLab Energy (Joules) }
    \end{subfigure}
    \begin{subfigure}[t]{0.31\textwidth}
    	\centering
    	\caption{Inter-Cloud Client Energy (Joules)}
        \includegraphics[keepaspectratio=true,width=58mm]{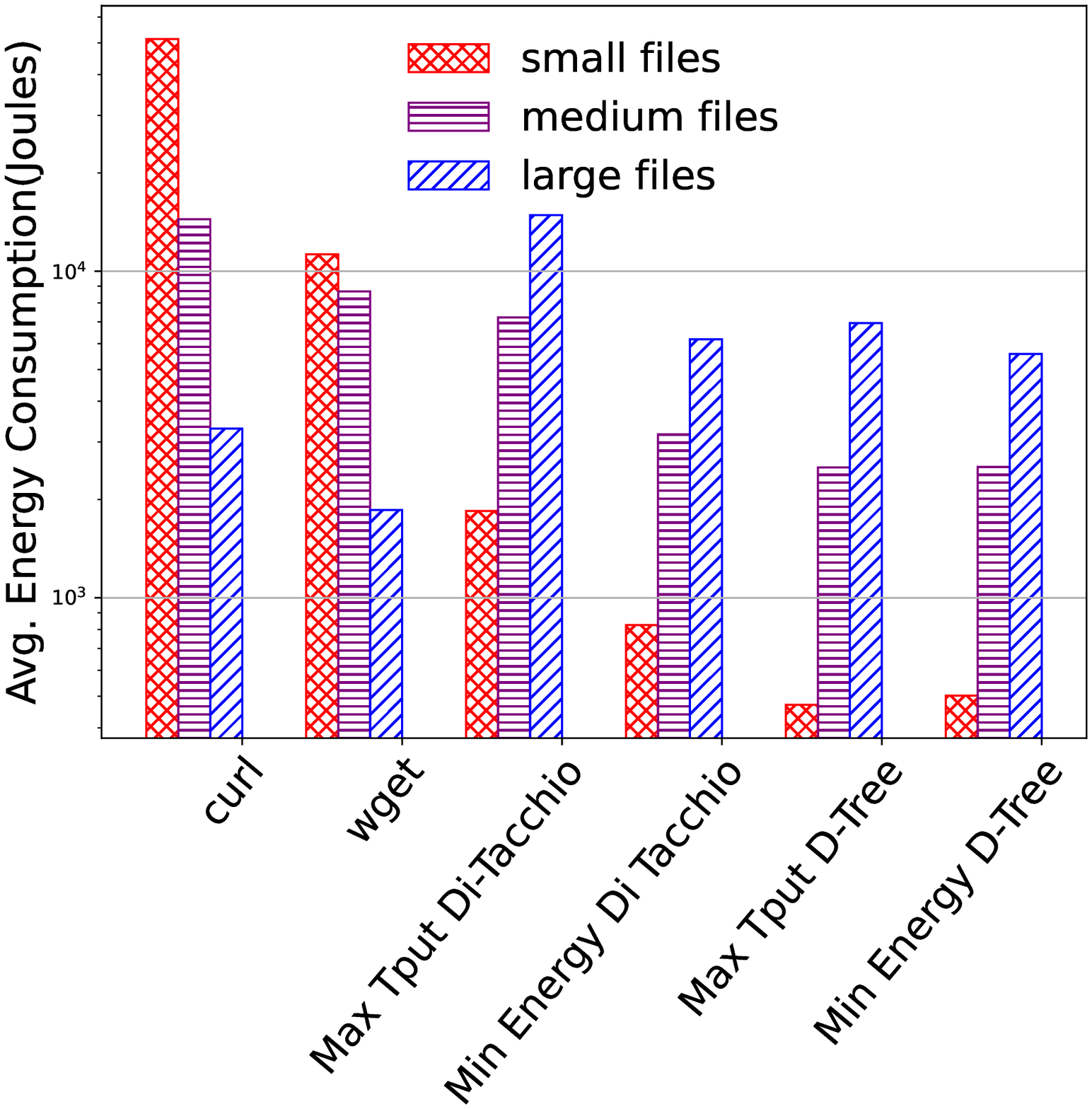}
        \vspace{-2mm}
        %\caption{Inter-Cloud Client Energy (Joules)}
    \end{subfigure}
     \caption{Achieved throughput (Mbps) and energy consumption (Joules) over 3 diverse testbeds. }
     \label{fig:throughputresults}
     \end{centering}
 \end{figure*}

\section{Experimental Evaluation}
We performed over 45,000 data transfers across three diverse wide-area network testbeds, collecting valuable experimental data encapsulating network conditions and characteristics. Testbeds utilized include (1) Chameleon Cloud ~\cite{keahey2020lessons}, server located at the University of Chicago and client located at the Texas Advanced Computing Center; (2) CloudLab ~\cite{Duplyakin+:ATC19}, server located at the University of Wisconsin and client located at the University of Utah; (3) Inter-Cloud, server located at the Texas Advanced Computing Center (part of Chameleon Cloud) and client located at the University of Utah (part of CloudLab). Both the Chameleon nodes run on either a Dell PowerEdge R630 containing 24 CPU cores distributed in dual-socket Intel Xeon E5-2670 v3 "Haswell" processors, each containing 12 cores, or on a PowerEdge R740 containing two Intel Xeon Skylake CPUs (each with 12 cores / 24 threads) ~\cite{keahey2020lessons}. The client within the CloudLab architecture runs on an HPE ProLiant XL170r server containing 10 CPU cores plus hyper-threading distributed in an Intel E5-2640v4 "Broadwell" processor having 64 GiB of RAM. The server within the CloudLab testbed and the client within the inter-cloud testbed run on Cisco's UCS SFF 220 M4 and UCS LFF 240 M4, respectively. Both contain 2 Intel E5-2630 "Haswell" processors, each having eight cores plus hyper-threading and 128 GiB of RAM ~\cite{Duplyakin+:ATC19}. The server within the Inter-Cloud testbed shares the same specifications as the Chameleon Cloud server. A specification overview is provided in Table III.

Experimental data transfers were performed during peak and non-peak hours, utilizing three diverse datasets containing different characteristics. These include: (1) the small file size dataset consisting of 20,000 HTML files derived from the common crawl project~\cite{CommonCrawl2020}; (2) the medium file size dataset consisting of 5,000 image files derived from Flickr~\cite{YahooFlickr}; (3) the large file size dataset consisting of 128 video files from Jiku~\cite{JikuVideoDataSet2020}. Complete dataset characteristics are specified in Table IV. 

On all of our testbeds, we measure the client’s energy consumption using Intel’s Running Average Power Limit (RAPL), which uses a software model to accurately estimate power consumption based on hardware performance counters and I/O models. David et al. ~\cite{david2010}  and Hahnel et al. ~\cite{ahnel2012}  highlighted RAPL’s precision in measuring both memory and CPU power consumption. To distinguish data transfer power consumption from total system power consumption, we subtracted the system baseline power consumption from the total power.
In order to fairly compare the efficiency of our dynamic minimum energy algorithm and maximum throughput algorithm with other transfer solutions, we utilize the extreme use cases. To achieve this, we use an SLA policy that informs our dynamic energy constraint algorithm to transfer data with the least amount of energy consumption by utilizing cross-layer optimal kernel-level and application-level parameters. We call this SLA policy the minimum energy decision tree (i.e., Min Energy D tree) SLA. To test our maximum throughput optimization algorithm, we use an SLA policy maximum throughput SLA (i.e., Max Throughput D tree) that enforces our algorithm to transfer data with the maximum throughput rate achievable, utilizing cross-layer optimal kernel-level and application-level parameters.

 \begin{table}[!t]
\centering
\label{offlineanalysistime}
\caption{Required time for offline analysis}
\rowcolors{2}{}{gray!10}
\begin{tabular}{|c|c|c|c|}
\hline
\begin{tabular}[c]{@{}c@{}}\textbf{TestBed}  \\  \textbf{(\#of logs)}\end{tabular}                                                  & \begin{tabular}[c]{@{}c@{}}\textbf{Cloudlab} \\ \textbf{(9841)}\end{tabular} & \begin{tabular}[c]{@{}c@{}}\textbf{Chameleon}\\ \textbf{(18325)}\end{tabular} & \begin{tabular}[c]{@{}c@{}}\textbf{InterCloud}\\ \textbf{(11228)}\end{tabular} \\ \hline \hline
\begin{tabular}[c]{@{}c@{}}Average Decision tree \\ Construction Time  \\  (Second)\end{tabular}                  & 4.5397                                                     & 10.41378                                                    & 4.3038                                                       \\ 
\begin{tabular}[c]{@{}c@{}}Hash Table \\ construction \\ time from tree\\ traversing time\\ (Second)\end{tabular} & 9.025355                                                   & 13.0955                                                     & 20.253139                                                    \\ \hline
\end{tabular}
\label{offlineanalysistime}
\end{table}

\begin{table}[!t]
\centering
%\caption{table description}
\label{t_sim_5}
\caption{Dataset Characteristics}
\rowcolors{2}{}{gray!10}
\begin{tabular}{|l|l|l|l|l|}
\hline 
\multicolumn{1}{|l|}{\textbf{Dataset}} & \multicolumn{1}{l|}{\textbf{Num Files}} & \multicolumn{1}{l|}{\textbf{Total Size}} & \multicolumn{1}{l|}{\textbf{Avg. File Size}} &
\multicolumn{1}{l|}{\textbf{Std. Dev.}} \\ 
\hline \hline
Small & 20,000 & 1.94 GB & 101.92 KB & \makecell{29.06 KB} \\  
Medium & 5,000 & 11.70 GB & 2.40 MB & \makecell{0.27 MB} \\
Large & 128  & 27.85 GB & 222.78 MB & \makecell{15.19 MB} \\
\hline
\end{tabular}
\end{table}

The experimental results in Figure 2 show how our proposed decision tree-based max throughput and min energy algorithms compare to Di Tacchio's algorithm implemented in~\cite{di2019cross}. Di Tacchio et al. developed real-time tuning heuristics to optimize throughput and minimize energy consumption by tuning both application-level data transfer parameters and kernel-layer parameters during HTTP transfers. Furthermore, we compared our algorithms to two common data transfer baseline tools: (1) curl, an open-source tool used to transfer data, and (2) wget, a free command-line tool used to retrieve files from the web. For a fair comparison between all algorithms and baseline tools, we utilized the datasets specified in table III when performing experimental data transfers. In addition, since the baseline tools did not support service-level agreements (SLAs), we set the SLAs of our models/algorithms to two diametrical cases: (1) maximum achievable throughput and (2) minimum achievable energy consumption. 

Figure 2 compares throughput performance and energy consumption across three diverse testbeds: (1) Chameleon, (2) CloudLab, and (3) Inter-Cloud. As anticipated, curl and wget produced sub-par results across all testbeds for all data transfers due to the absence of parameter optimization. Di Tacchio et al.'s algorithms performed better than all the baseline tools. As demonstrated in sub-figures, 2(a) and 2(d), our algorithm outperforms all other algorithms in both throughput performance and energy consumption cases within a high bandwidth and high Bandwidth Delay Product (BDP) network environment (Chameleon Cloud). Di Tacchio's heuristic does not take advantage of past data transfer history and must adjust parameters strictly on fluctuating network conditions. 

Figures 2(b) and 2(e) show how our dynamic max throughput and min energy algorithm outperforms all other algorithms in lower bandwidth and lower BDP network environments (Cloudlab testbed). Sub-figures 2(c) and 2(f) demonstrate how our algorithms outperform all other algorithms and baseline tools for datasets containing small size HTML files, datasets containing medium size image files, and a dataset containing large video files in the intercloud testbed. Decision tree max throughput outperformed Max throughput (Di Tacchio) across all datasets in throughput improvement of HTML data transfers by 180\%, image data transfers by 120\%, and video data transfers by 52\%.   Additionally, the decision tree minimum energy algorithm decreased energy consumption by an additional 37\% for HTML data transfers, 23\% for image data transfers, and 6\% for video data transfers, with respect to the minimum Energy(Di Tacchio) algorithm. The cost of overestimating or underestimating parameter values and compute resources are expensive. Overestimating parameter values and compute resources can increase energy consumption. 

\begin{figure}[]
    \begin{centering}
    \includegraphics[keepaspectratio=true,width=75mm]{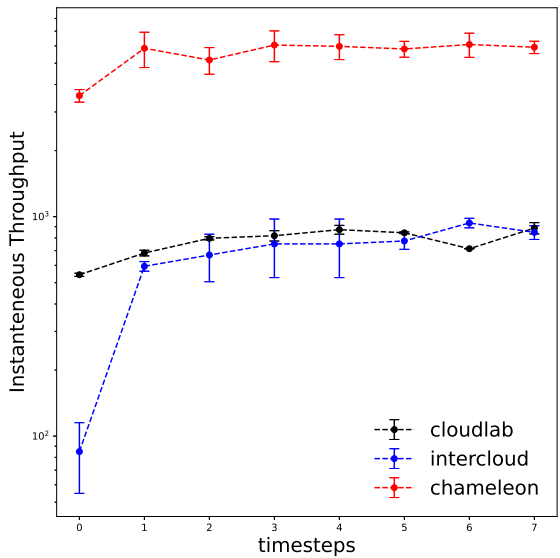}
    % \vspace{-2mm}
        %\caption{Chameleon Throughput (Mbps)}
    \caption{The convergence of achieved throughput with decision tree max throughput algorithm for all three testbeds. The x axis shows the timesteps and y axis is instantaneous throughput in log scale.}
     \vspace{-3mm}
     \label{fig:convergence}
     \end{centering}
 \end{figure}

Conversely, underestimating parameter values and compute resources can degrade throughput performance. Utilizing decision-search-tree based clustering techniques on past data transfer history logs allows our algorithms to accurately estimate near-optimal data transfer parameters for a given SLA based on the current network conditions. This causes our algorithms to converge faster to optimal parameters. Our dynamic maximum throughput and min energy algorithms converge quickly to optimal cross-layer parameter values based on real-time network feedback and historical log analysis. As shown in figure~\ref{fig:convergence}, with the decision tree approach, the convergence to optimal throughput takes the minimal number of timesteps\footnote{For each data point in the line, it denotes the mean of all the instantaneous throughput at that timestep and the vertical bars at each data point shows the distribution of all the instantaneous throughput at that timestep. From the plot it could be deduced that decision tree throughput algorithm converges to the maximum achievable throughput within one timestep.}. This faster convergence is because the decision tree combines the current network condition and knowledge from offline historical analysis to provide faster convergence towards maximally achievable throughput.

\section{Related Work}
Improving data transfer throughput in wide-area high-speed networks is a challenging task that depends on many dynamic factors, including but not limited to: network conditions, network settings, data transfer parameter configuration, and dataset characteristics. Therefore, it is difficult to develop predictive models to accurately capture hidden throughput patterns in fluctuating network environments. Several works on application-level parameter tuning during data transfer mostly propose non-scalable and static solutions to the problem with some predefined values for the subset of the problem space~\cite{deelman2006},\cite{R_Hacker02},\cite{kola2004},\cite{R_Dinda05}.%
 Earlier research focused on developing predictive models based on three methodologies: analytical, empirical, and model-free~\cite{R_Dinda05},%~\cite{Altman_2006},
~\cite{balman2009dynamic}, ~\cite{Yin_2011}. Hacker et al. \cite{R_Hacker02} developed an analytical model correlating throughput with various network parameters. Arslan et al. ~\cite{arslan2014locality}
applied similar optimization tehcniques to achieve locality and network-aware reduce task scheduling for data-intensive applications. Yildirim et al. \cite{Yildirim_2011} developed an empirical approach by applying Newton's method to find the optimal number of TCP streams via parallelism that will maximize throughput. Kim et al.~\cite{kim2015highly} considered the combined effect of parallelism, pipelining, and concurrency on end-to-end data transfer throughput. In our prior work \cite{di2019cross} and \cite{Rodolph}, we developed a combination of model-free and empirical algorithms to maximize throughput by utilizing both machine learning techniques and online dynamic parameter tuning. This paper extends our prior work by using a novel decision tree-based model capable of dealing with the aleatoric uncertainty presented in the historical logs.

%%%%%%%%%%%%%%%%%%%%%%%%%%%%%%%%%%%%%%%%%%%%%%%%%%%%%%%%%%%%%%%%%%%%%%%%%%%%%%%%%
\section{Conclusion}
Augmenting data transfer throughput and minimizing energy consumption during data transfer in end hosts over high-speed, long-distance networks is becoming progressively challenging. Numerous factors such as fluctuating network conditions, limitations of underlying transfer protocols, dataset characteristics, network characteristics, and data transfer parameter settings must be considered to achieve optimal or close to optimal performance. A multifaceted approach to improving data transfer throughput and minimum energy consumption involves optimally and dynamically adjusting application-layer and kernel-layer transfer parameters based on real-time network conditions and dataset characteristics. In this paper, we presented a novel two-phase dynamic throughput predictive optimization model that utilizes offline decision-search-tree based learning techniques to encapsulate and categorize historical data transfer log information and an online search optimization algorithm to find optimal transfer parameters. Furthermore, we also explore ensemble methods to tackle the uncertainty while carrying out the offline phase of analyzing historical log data and building a decision search tree. The experimental evaluation shows that our decision tree-based model outperforms state-of-the-art solutions in this area by achieving 117\% higher throughput on average while consuming 19\% less energy at the end systems during active data transfers.

 %During the offline analysis phase, we constructed decision search trees with uncertainty quantification based on historical data transfer log files encapsulating network characteristics, dataset characteristics, and external network conditions. During the online phase, we utilized novel search tree optimization algorithms to obtain optimal transfer parameters based on real-time network conditions and dataset characteristics to optimize data transfer throughput. 

\section*{Acknowledgements}
This project is in part sponsored by the National Science Foundation (NSF) under award numbers CCF-2007829, OAC-1842054, OAC-1724898. We also would like to thank the Chameleon Cloud and CloudLab for letting us use their resources in our experiments.

\vspace{-1mm}
\bibliographystyle{IEEEtran}
\bibliography{references}

\end{document}